\documentclass[aps,a4paper, preprint, superscriptaddress,preprintnumbers,floatfix,nofootinbib,amsmath,amssymb]{revtex4}
\usepackage{url}
\usepackage{hyperref}
\usepackage{cleveref}

\usepackage{amstext,amssymb}
\usepackage{amsmath}
\usepackage{graphicx}
\usepackage{xspace}
\usepackage{color}
\usepackage{units}

\usepackage{feynmp}
\usepackage{cancel}
\usepackage{bm}
\usepackage{slashed} 
\usepackage{multirow}
\newcommand{\be}{\begin{equation}}
\newcommand{\ee}{\end{equation}}
\newcommand{\bea}{\begin{eqnarray}}
\newcommand{\eea}{\end{eqnarray}}
\newcommand{\nn}{\nonumber}

\def\s1{\hat s}

\usepackage{slashed}

\newcommand{\nua}[1]{\ensuremath{\rlap{\kern-2.5pt\ensuremath{\overset{\scriptscriptstyle(-)}{\phantom{\nu}}}}{\ensuremath{{\nu}_{#1}}}}\xspace}

\begin{document}
\title{A modular $A_4$  symmetric  Scotogenic model for Neutrino mass and Dark Matter}
\author{ Mitesh Kumar Behera}
\email{miteshbehera1304@gmail.com}
\affiliation{School of Physics,  University of Hyderabad, Hyderabad - 500046,  India}
\author{ Shivaramakrishna Singirala}
\email{krishnas542@gmail.com}
\affiliation{School of Physics,  University of Hyderabad, Hyderabad - 500046,  India}
\author{Subhasmita Mishra}
\email{subhasmita.mishra92@gmail.com}
\affiliation{Department of Physics, IIT Hyderabad, Kandi - 502285, India}
\author{ Rukmani Mohanta}
\email{rmsp@uohyd.ac.in}
\affiliation{School of Physics,  University of Hyderabad, Hyderabad - 500046,  India}

\begin{abstract}
Modular symmetries have been impeccable in neutrino and quark sectors.  This motivated us, therefore, to propose a variant of scotogenic model based on modular $A_4$ symmetry to realize the neutrino mass generation at one-loop level through radiative mechanism.  Alongside, the lepton flavour violating process $\mu \to e \gamma$ and the  muon $g-2$ anomaly  are also  addressed. The lightest Majorana fermions turn out to be potential dark matter candidates, made stable by suitable assignment of modular weights. The relic density of the same has been computed with annihilations mediated by inert scalars and new $U(1)$ gauge boson. 
\end{abstract}

\maketitle
\flushbottom

\section{INTRODUCTION}
\label{sec:intro}
Various experimental observations over the last few decades have conclusively established the robustness of the Standard Model (SM). Nonetheless, there are a few issues demonstrating the  presence of physics beyond the SM, for example, the nature and existence of dark matter (DM) \cite{Zwicky:1937zza, Rubin:1970zza, Clowe:2003tk,Bertone:2004pz,ArkaniHamed:2008qn,Dodelson:1993je},  small but non-vanishing  neutrino masses \cite{RoyChoudhury:2019hls,Fukuda:1998mi,Aghanim:2018eyx},  observed baryon asymmetry of the Universe  \cite{Sakharov:1967dj,Kolb:1979qa, Davidson:2008bu,Buchmuller:2004nz,Strumia:2006qk}, origin of flavor structure, etc. Therefore,  apprehending the nature of physics beyond the standard model (BSM) gets inescapable, and in this context, symmetry is  assumed to play a significant role, e.g., ensuring the appropriate mechanism for achieving the tiny neutrino masses, 
stability of DM, confining flavour structure, and so on. It is thus, intriguing to build  models beyond the SM adopting  new symmetries.
\vspace{2mm}

The Scotogenic model, proposed by Ma \cite{Ma:2006km,Ma:2009gu} is probably  the simplest model that generates the small neutrino masses at one-loop level  and also simultaneously accounts for the dark matter (both inert scalar and fermionic), see for example a legion of works in the literature \cite{LopezHonorez:2006gr,Gustafsson:2007pc,Dolle:2009fn, Suematsu:2009ww, Schmidt:2012yg, Singirala:2016kam} and references therein. Various other works have realized neutrino mass at one-loop \cite{Restrepo:2019ilz,Babu:2019mfe,Chen:2019okl,Ma:2019yfo,Nomura:2019lnr}. Further, the pioneering work of introducing modular flavor symmetries to quark and neutrino sectors is seen in the literature of  \cite{Feruglio:2017ieh,Feruglio:2017spp,King:2020qaj} to highlight predictable flavor structures. The basic idea behind using the modular symmetry  is either to nullify or minimize the necessity to include flavon fields other than modulus $\tau$. Some of the effective models based on modular symmetry of recently published papers \cite{King:2017guk,Altarelli:2010gt,Ishimori:2010au,King:2015aea} justify the statement above. The  breaking of  flavor symmetry takes place when this complex modulus $\tau$ acquires VEV. The main issue of the perplexing  vacuum alignment  is avoided, the only requirement is a certain kind of mechanism which can fix the modulus $\tau$. Resultantly, this has prompted a restoration of the possibility that modular symmetries are symmetries of the extra dimensional space-time with Yukawa couplings dictated by their modular weights \cite{Criado:2018thu} hence, transform systematically under this framework, where there is a functional dependence of these couplings on modular forms, which verily are holomorphic function of $\tau$. To put it in a different way, these couplings come to pass under a non-trivial representation of a non-Abelian discrete flavor symmetry approach \cite{Altarelli:2010gt}, to such an extent that it can remunerate the utilization of flavon fields, which undoubtedly are not required or limited in understanding the flavor structure. In reference to above,  it was fathomed that there are numerous groups accessible i.e., basis characterized under modular group of $A_4$ \cite{Abbas:2020qzc,King:2020qaj,Wang:2019xbo,Lu:2019vgm, Kobayashi:2019gtp,Nomura:2019xsb,Behera:2020sfe}, $S_4$ \cite{Penedo:2018nmg,Gui-JunDing:2019wap,Liu:2020akv,Kobayashi:2019xvz,Wang:2019ovr}, $A_5$ \citep{Ding:2019zxk,Novichkov:2018nkm}, larger groups \cite{Kobayashi:2018wkl}, various other modular symmetries and double covering of $A_4$ \cite{Nomura:2019lnr,Ma:2015fpa,Mishra:2019oqq,Novichkov:2020eep}, predictions regarding masses, mixing \cite{King:2013xba,King:2009fk}, and CP phases distinctive to quarks and/or leptons are made. \vspace{1mm}

This paper contains, minimal scotogenic model~\cite{Avila:2019hhv,Dasgupta:2019rmf,Ma:2012ez,Bouchand:2012dx,Fraser:2015mhb,Rojas:2018wym,Hagedorn:2018spx,Kitabayashi:2018bye,Pramanick:2019qpg,Tang:2017rhv}, constructed, based on modular $A_4$ symmetry in which mass generation for neutrinos is done at one-loop level and it also provides a stable DM candidate.  A minimal Scotogenic model can be appreciated by using modular forms having higher weights, which have a dependence on weight-2 triplet Yukawa couplings. Thus, field contents and model's structure are much simpler than previous models~\cite{Nomura:2019jxj,Okada:2019mjf}. Our model encompasses two different sets of SM singlet heavy neutrinos i.e., $N_{Ri}$ \& $S_{Li}$,  $(i=1,2,3)$, which transform  as  triplets under $A_4$, with modular weight $k_I=-1$ and $k_I=1$ respectively. Likewise, the inert Higgs doublet is allocated a non-zero modular weight as $k_I=-2$.  Interestingly, modular weights help in impersonating the additional $Z_2$ symmetry, hence, it is not necessary to use $Z_2$ symmetry for constructing scotogenic model and realizing the stability of DM.

The layout of this paper is as follows. In Sec. \ref{sec:realization} we outline our model framework  with discrete $A_4$ modular flavor symmetry and its appealing feature resulting in simple mass structure for the charged  and neutral leptons with two types of sterile neutrinos.  We then provide a brief discussion on the generation of  light neutrino masses and their mixing in  Sec. \ref{sec:radiative}.  In Sec. \ref{sec:analysis} numerical correlational study between observables of neutrino sector and input model parameters is established. Comments on lepton flavour violating decays $\mu \to e \gamma$ decays and muon $g-2$ anomalies are presented  in Sec.\ref{sec:comment}. Further, Sec.
 \ref{sec:dark} comprises the discussion on fermionic dark matter followed  by our conclusions in  Sec.\ref{sec:con}.

\section{MODEL FRAMEWORK}
\label{sec:realization}
Here, we take the privilege of introducing the model framework,  investigating the impact of $A_4$ modular symmetry on neutrino and dark matter phenomenology. The SM particle spectrum is enriched with three right-handed ($N_R$) and three left-handed heavy fermions ($S_L$) to meet the purpose. We impose a local $U(1)_X$ symmetry to avoid certain unwanted interactions and a scalar singlet $\rho$ to break it spontaneously. The scalar sector is extended with an inert scalar doublet $\eta$, to realize neutrino mass at one-loop. The assigned modular weight mimics $Z_2$ symmetry by playing a vital role in forbidding the neutrino mass at tree-level and also in stabilizing the fermionic dark matter. The representation of different fields of the model under ${ SU(2)_L\times U(1)_Y \times U(1)_X} \times A_4$ symmetries and their modular weights are given in the Table~\ref{tab:fields-linear}. In addition, the non-trivial transformation of Yukawa and scalar couplings and their modular weights are furnished in Table~\ref{tab:couplings}.
 
\begin{table}[h!]
\begin{center}
\begin{tabular}{|c||c|c|c|c|c|c||c|c|c|c|c|}\hline\hline  
  & \multicolumn{6}{c||}{Fermions} & \multicolumn{3}{c|}{Scalars}\\ \hline \hline

& ~$e_R$~& ~$\mu_R$~  & ~$\tau_R$~& ~$L_L$~& ~$N_R$~& ~$S_L$~& ~$H$~&~ $\eta$ ~&~$\rho$ \\ \hline 
$SU(2)_L$ & $1$  & $1$  & $1$  & $2$  & $1$  & $1$  & $2$ &$2$   & $1$     \\\hline 
$U(1)_Y$   & $-1$ & $-1$ & $-1$ & $-\frac12$  & $0$ & $0$  & $\frac{1}{2}$ & $\frac{1}{2}$  & $0$   \\\hline
$U(1)_X$   & $1$ & $1$ & $1$   &$1$  & $1$  & $0$  & $0$ &$0$ &$1$  \\\hline
$A_4$ & $1$ & $1'$ & $1''$ & $1, 1^{\prime \prime}, 1^{\prime }$ & $3$ & $3$ & $1$ & $1$  &$1$ \\ \hline
$k_I$ & $-1$ & $-1$ & $-1$ & $-1$ & $-1$ & $1$ & $0$  & $-2$ & $0$\\ \hline

\hline
\end{tabular}
\caption{Particle content of the model and their charges under ${ SU(2)_L\times U(1)_Y\times U(1)_X}\times A_4 $, where $k_I$ is the modular weight.}
\label{tab:fields-linear}
\end{center}
\end{table}

\begin{table}[h!]
\begin{center}
\begin{tabular}{|c||c|c|c|c|c|}\hline
{Couplings}  & ~{ $A_4$}~& ~$k_I$~     \\\hline 
{ $\bm{Y}=(y_1,~y_2,~y_3)$} & ${\bf 3}$ & ${\bf 2}$      \\\hline
$\bm{\lambda_\eta}$  & $\bf{1}$ & $\bf{8}$ \\ \hline
$\bm{\lambda^\prime_{\eta }}$  & $\bf{1}$ & $\bf{4}$ \\ \hline 
\end{tabular}
\caption{Transformation of the Yukawa and quartic couplings under $A_4$ symmetry and their corresponding modular weights.}
\label{tab:couplings}
\end{center}
\end{table}

The scalar potential of the model is given by  
\begin{eqnarray}
\mathcal{L}_V &=& \mu^2_H (H^\dagger H)+\lambda_H (H^\dagger H)^2+\mu^2_{\rho}(\rho^\dagger\rho)+\lambda_{\rho}(\rho^\dagger\rho)^2+\lambda_{H\rho}(H^\dagger H)(\rho^\dagger\rho) +\lambda_\eta \zeta_2(\eta^\dagger \eta)^2 \nn \\
&& +\lambda^\prime_{\eta}\Big[\mu^2_{\eta}(\eta^\dagger \eta)+\zeta_3(H^\dagger H)(\eta^\dagger \eta) +\zeta_4(H^\dagger \eta)(\eta^\dagger H)+ \frac{\zeta_5}{2}((H^\dagger \eta)^2 +~{\rm H.c}) \nn\\
&&  + \zeta_6(\eta^\dagger \eta)(\rho^\dagger\rho)\Big].
\end{eqnarray}

Here, $H = \left(0 ~~(v+h)/\sqrt{2}\right)^T$ is the SM Higgs doublet, $\eta = \left(\eta^+ ~~(\eta_R+i \eta_I)/\sqrt{2}\right)^T$ denotes the inert doublet and the complex scalar $\rho = \frac{1}{\sqrt{2}}(v_{\rho}+h_{\rho}+iA_{\rho})$ breaks the $U(1)_X$  local gauge symmetry spontaneously. The mass mode of $A_\rho$ is eaten up by the $U(1)_X$ associated gauge boson $Z^\prime$, attains the mass $M_{Z^\prime} = g_X v_\rho$. In the above potential, $\zeta_i$'s are the free parameters and the scalar coupling $\lambda^\prime_{\eta}$ is the singlet representation of $A_4$ with modular weight 4, which can be expressed in terms of the components  of weight-2  triplet Yukawa couplings \cite{Feruglio:2017spp},
\begin{eqnarray}
\lambda^\prime_{\eta}=y_1^2 +2y_2y_3.
\end{eqnarray}
For simplicity, we avoid $H-\rho$ mixing i.e., $\lambda_{H\rho}=0$. The mass spectrum of scalar sector \cite{Lindner:2016kqk} can be written as follows:
\begin{eqnarray}
&&M_{h}^2 = 2\lambda_H v^2,\nn\\
&&M_{\rho}^2 = 2\lambda_\rho v_\rho^2,\nn\\
&&M_{\eta^\pm}^2 = \lambda^\prime_{\eta} \left[ \mu^2_{\eta} + \zeta_3 \frac{v^2}{2} + \zeta_6 \frac{v_\rho^2}{2} \right],\nn\\
&&M_{\eta_R,\eta_I}^2 = \lambda^\prime_{\eta} \left[\mu^2_{\eta} + (\zeta_3+\zeta_4\pm \zeta_5) \frac{v^2}{2} + \zeta_6 \frac{v_\rho^2}{2}\right].
\end{eqnarray}

In order to construct a simplified version of charged leptons mass matrix, left-handed doublets (i.e., three generations ($L_{e_L}, L_{\mu_L}, L_{\tau_L} $)) are considered to transform as $\bm{1}, \bm{1}^{\prime\prime}, \bm{1}^{\prime}$ respectively under the $A_4$ symmetry with assignment of modular weight, $k_I=-1$ for each generation. Analogously, the right-handed charged leptons ($e_R,\mu_R,\tau_R$) transform  under $A_4$ as  $\bm{1}, \bm{1}^{\prime}, \bm{1}^{\prime\prime}$, and carry a modular weight, $k_I=-1$.  
The SM Higgs is uncharged under the new symmetries,  to make the scenario a bit simplistic. 

The charged leptons interaction Lagrangian is given by
\begin{align}
 \mathcal{L}_{M_\ell}  
                   &= y_{\ell_{}}^{ee}  \overline{L}_{e_L} H e_R +  y_{\ell_{}}^{\mu \mu}  \overline{L}_{\mu_L} H \mu_R +  y_{\ell_{}}^{\tau \tau}  \overline{L}_{\tau_L} H \tau_R
                    + {\rm H.c.}. \label{Eq:yuk-Mell} 
\end{align}
The mass matrix for charged leptons achieves a diagonal structure, following, the spontaneous breaking of electroweak gauge symmetry. Moreover, one can obtain the observed masses for the charged leptons by adjusting the Yukawa couplings. Hence, the obtained mass matrix is represented as follows
\begin{align}
M_\ell = \begin{pmatrix}  y_{\ell_{}}^{ee} v/\sqrt{2}  &  0 &  0 \\
                                       0  &  y_{\ell_{}}^{\mu \mu} v/\sqrt{2}  &  0 \\
                                       0  &  0  &  y_{\ell_{}}^{\tau \tau} v/\sqrt{2}        \end{pmatrix}  =
                     \begin{pmatrix}  m_e  &  0 &  0 \\
                                       0  &  m_\mu  &  0 \\
                                       0  &  0  &  m_\tau      \end{pmatrix},                
\label{Eq:Mell} 
\end{align}
where $m_e$, $m_\mu$ and $m_\tau$ are the observed charged lepton masses.

\subsection{Dirac and pseudo-Dirac interaction terms for the  neutrinos}
The right (left) handed heavy fermions contrary to SM leptons are considered as triplet under $A_4$ modular group with a $U(1)_X$ charge of $1 (0)$ and modular weight $k_I=-1 (+1)$. The usual Dirac interactions of neutrinos with SM Higgs can not be defined with aforesaid charges. The introduction of modular Yukawa couplings with transformation represented in Table\,\ref{tab:couplings} along with inert scalar doublet $\eta$ are necessary to write such interactions. Moreover, the Yukawa couplings {$\bm{Y}(\tau) = \left(y_{1}(\tau),y_{2}(\tau),y_{3}(\tau)\right)$}, are expressed in terms of Dedekind eta-function  $\eta(\tau)$ and its derivative, as discussed in  (Appendix of \citep{Feruglio:2017spp}).
Hence, the invariant Dirac interaction Lagrangian, which involves the active neutrinos along with the right and left-handed heavy fermions, can be represented  in the following  forms:
\begin{align}
 \mathcal{L}_{D}  
                   &= \alpha_D   \overline{L}_{e_L} \widetilde{\eta} (\bm{Y} N_R)_{1}   + \beta_D   \overline{L}_{\mu_L} \widetilde{\eta} (\bm{Y} N_R)_{1^{\prime}}
                   + \gamma_D   \overline{L}_{\tau_L} \widetilde{\eta} (\bm{Y} N_R)_{1^{\prime \prime}}                       + {\rm H.c.}, \label{Eq:yuk-MD} 
\end{align}
\begin{align}
 \mathcal{L}_{LS}  
                   &= \left[\alpha'_D   \overline{L}_{e_L} \widetilde{\eta} (\bm{Y} S_L^c)_{1}   + \beta'_D   \overline{L}_{\mu_L} \widetilde{\eta} (\bm{Y} S_L^c)_{1^{\prime}}
                   + \gamma'_D   \overline{L}_{\tau_L} \widetilde{\eta} (\bm{Y} S_L^c)_{1^{\prime \prime}}\right] \frac{\rho}{\Lambda} + {\rm H.c.}. \label{Eq:yuk-LS} 
\end{align}
Adjacently, the $A_4$ and $U(1)_X$ symmetric charges for heavy fermions are imposed in such a way that their usual Majorana mass terms are forbidden. However, the mixing between the additional leptons are allowed, which can be written as follows \cite{Behera:2020sfe}
\begin{eqnarray}
 \mathcal{L}_{M_{RS}}  
                   &=& \left[\alpha_{NS} \bm{Y} (\overline{S_L} N_R)_{\rm symm} + \beta_{NS} \bm{Y} (\overline{S_L} N_R)_{\rm Anti-symm} \right]\rho^\dagger   + {\rm H.c.} \nn \\
                   &=&\alpha_{NS}\big[ y_1(2  \bar S_{L_1} N_{R_1} - \bar S_{L_2} N_{R_3} - \bar S_{L_3} N_{R_2})+y_2(2  \bar S_{L_2} N_{R_2} - \bar S_{L_1} N_{R_3} - \bar S_{L_3} N_{R_1}) \nn \\
&&+ y_3(2  \bar S_{L_3} N_{R_3} - \bar S_{L_1} N_{R_2} - \bar S_{L_2} N_{R_1}) \big] \rho^\dagger +\beta_{NS}\big[ y_1(  \bar S_{L_2} N_{R_3} - \bar S_{L_3} N_{R_2}) \nn\\ &&+y_2(  \bar S_{L_3} N_{R_1} - \bar S_{L_1} N_{R_3})+ y_3(  \bar S_{L_1} N_{R_2} - \bar S_{L_2} N_{R_1})\big] \rho^\dagger  + {\rm H.c.}                   \label{Eq:yuk-M} 
\end{eqnarray}

Here, $\alpha_{NS}$ and $\beta_{NS}$ represent free parameters, the first term in (\ref{Eq:yuk-M}) is symmetric and second term is anti-symmetric product for $\bar S_L N_R$ making $\bm{3_s}$ and $\bm{3_a}$ representations of $A_4$. Using $\langle \rho \rangle = v_\rho/\sqrt{2}$,  the resulting mass matrix is found to be
\begin{align}
M_{RS}&=\frac{v_\rho}{\sqrt2}
 \left(
 \frac{\alpha_{NS}}{3}\left[\begin{array}{ccc}
2y_1 & -y_3 & -y_2 \\ 
-y_3 & 2y_2 & -y_1 \\ 
-y_2 & -y_1 & 2y_3 \\ 
\end{array}\right]
+
\beta_{NS}
\left[\begin{array}{ccc}
0 &y_3 & -y_2 \\ 
-y_3 & 0 & y_1 \\ 
y_2 & -y_1 &0 \\ 
\end{array}\right]
\right).
\label{MRS_matrix}
\end{align}
The mass matrix for the six heavy leptons, in the basis $( N_R, S_L)^T$,  can be given as
\begin{eqnarray}
M_{Hf}= \begin{pmatrix}
0 & M_{RS}\\
M^T_{RS} & 0
\end{pmatrix},\label{mrs matrix}
\end{eqnarray}
which  upon diagonalization provides three doubly degenerate mass pairs ($M_k$) and the digonalization of $M_{RS}$ with a simplified form is discussed in \cite{Behera:2020sfe}.
\section{Radiative Neutrino mass}
\label{sec:radiative}
\begin{figure}[h!]
\begin{center}
\includegraphics[height=45mm,width=70mm]{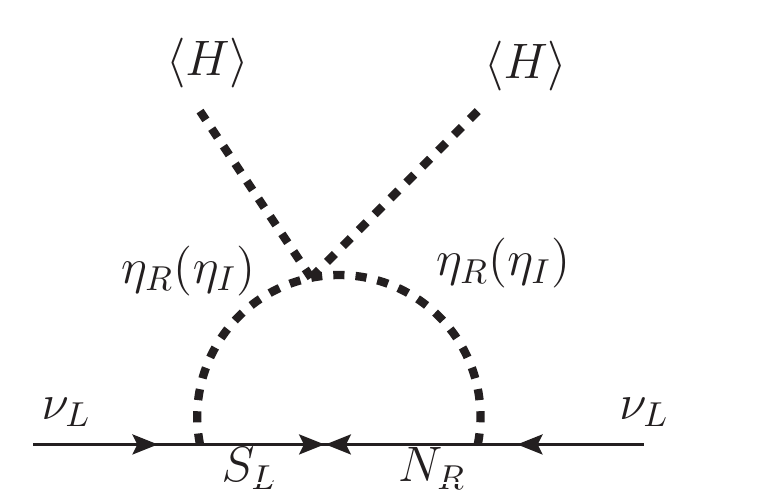}
\caption{Radiatively generated neutrino mass.}  \label{fig:radia}
\end{center} 
\end{figure}
Since, the usual Dirac mass terms of neutrinos with SM Higgs are forbidden by the assigned symmetries, one can  generate light neutrino masses  at one-loop level and the corresponding Feynman diagram is displayed in Fig \ref{fig:radia}. 

The expression of the neutrino mass from one loop radiative corrections is written as

\begin{equation}
(\mathcal{M_\nu})_{ij} = \sum_k\frac{{(Y_D)}_{ik}{(Y_{LS})}_{jk}}{16\pi^2} ~\left[\frac{M^2_{\eta_R}}{M^2_{\eta_R} - M_k^2}~ {\rm ln} \frac{M^2_{\eta_R}}{M_k^2}-\frac{M^2_{\eta_I}}{M^2_{\eta_I} - M_k^2}~ {\rm ln} \frac{M^2_{\eta_I}}{M_k^2}\right].                     \label{Eq:nmasss}
\end{equation}
Here, $M_k$ is the mass of the heavy fermion  inside the loop,  $Y_D$ and $Y_{LS}$ are the Yukawa coupling matrices correspond to the interaction of neutrinos with $N_R$ and $S_L$ respectively and are given by
\begin{align}
Y_D&=
\left[\begin{array}{ccc}
\alpha_D & 0 & 0 \\ 
0 & \beta_D & 0 \\ 
0 & 0 & \gamma_D \\ 
\end{array}\right]
\left[\begin{array}{ccc}
y_1 &y_3 &y_2 \\ 
y_2 &y_1 &y_3 \\ 
y_3 &y_2 &y_1 \\ 
\end{array}\right]_{LR}.
\label{Eq:MD}  
\end{align}

\begin{align}
Y_{LS}&= \frac{v_\rho}{\Lambda \sqrt{2}}
\left[\begin{array}{ccc}
\alpha^\prime_D & 0 & 0 \\ 
0 & \beta^\prime_D & 0 \\ 
0 & 0 & \gamma^\prime_D \\ 
\end{array}\right]
\left[\begin{array}{ccc}
y_1 &y_3 &y_2 \\ 
y_2 &y_1 &y_3 \\ 
y_3 &y_2 &y_1 \\ 
\end{array}\right]_{LS}.                   
\label{Eq:Mls} 
\end{align}
The mass matrix in eqn.\eqref{Eq:nmasss}, can be reduced to the simplified form as follows with the assumption $M^2_k \ll m^2_0$, where $m^2_0 = (M^2_{\eta_R} + M^2_{\eta_I})/{2}$.
\begin{eqnarray}
({\cal M}_\nu)_{ij}= \frac{\zeta_5 \lambda^\prime_{\eta} v^2}{16 \pi^2 m^2_0} \sum_k (Y_D)_{ik} (Y_{LS})_{kj} M_k,\label{Eq:nmasss2}
\end{eqnarray}
where, we have used $M^2_{\eta_R} - M^2_{\eta_I} =  \zeta_5 \lambda_{\eta}^\prime v^2$. When specific mass ranges are considered for $m_{\eta_R}$, $m_{\eta_I}$ and $M_k$, this formula helps to generate both linear seesaw and inverse seesaw \cite{Deppisch:2004fa,Dev:2012sg,Hirsch:2009mx}. The neutrino mass matrix  (\ref{Eq:nmasss2}) is numerically  diagonalized through the relation $U^\dagger  \mathbb{M} U = {\rm diag}(m_1^2, m_2^2,m_3^2)$, where $\mathbb{M} = {\cal M}_\nu {\cal M}_\nu^\dagger$ and $U$ is an unitary matrix.
Thus, the neutrino mixing angles can be extracted from the matrix elements of the  diagonalizing matrix $U$, through  the generic expressions:
\begin{eqnarray}
\sin^2 \theta_{13}= |U_{13}|^2,~~~~\sin^2 \theta_{12}= \frac{|U_{12}|^2}{1-|U_{13}|^2},~~~~~\sin^2 \theta_{23}= \frac{|U_{23}|^2}{1-|U_{13}|^2}. \label{eq:UPMNS}
\end{eqnarray}
Next, we attempt to determine the  Jarlskog invariant ($J_{CP}$) as well as the effective Majorana mass parameter ($\langle m_{ee}\rangle$) through the following relations: 
\begin{eqnarray}
&&J_{CP} = \text{Im} [U_{e1} U_{\mu 2} U_{e 2}^* U_{\mu 1}^*] = s_{23} c_{23} s_{12} c_{12} s_{13} c^2_{13} \sin \delta_{CP}. \\
&& \langle m_{ee}\rangle=|m_{\nu_1} \cos^2\theta_{12} \cos^2\theta_{13}+ m_{\nu_2} \sin^2\theta_{12} \cos^2\theta_{13}e^{i\alpha_{21}}+  m_{\nu_3} \sin^2\theta_{13}e^{i(\alpha_{31}-2\delta_{CP})}|.\nn \\
\end{eqnarray}
\section{Numerical Analysis}
\label{sec:analysis}
\begin{figure}[h!]
\begin{center}
\includegraphics[height=50mm,width=75mm]{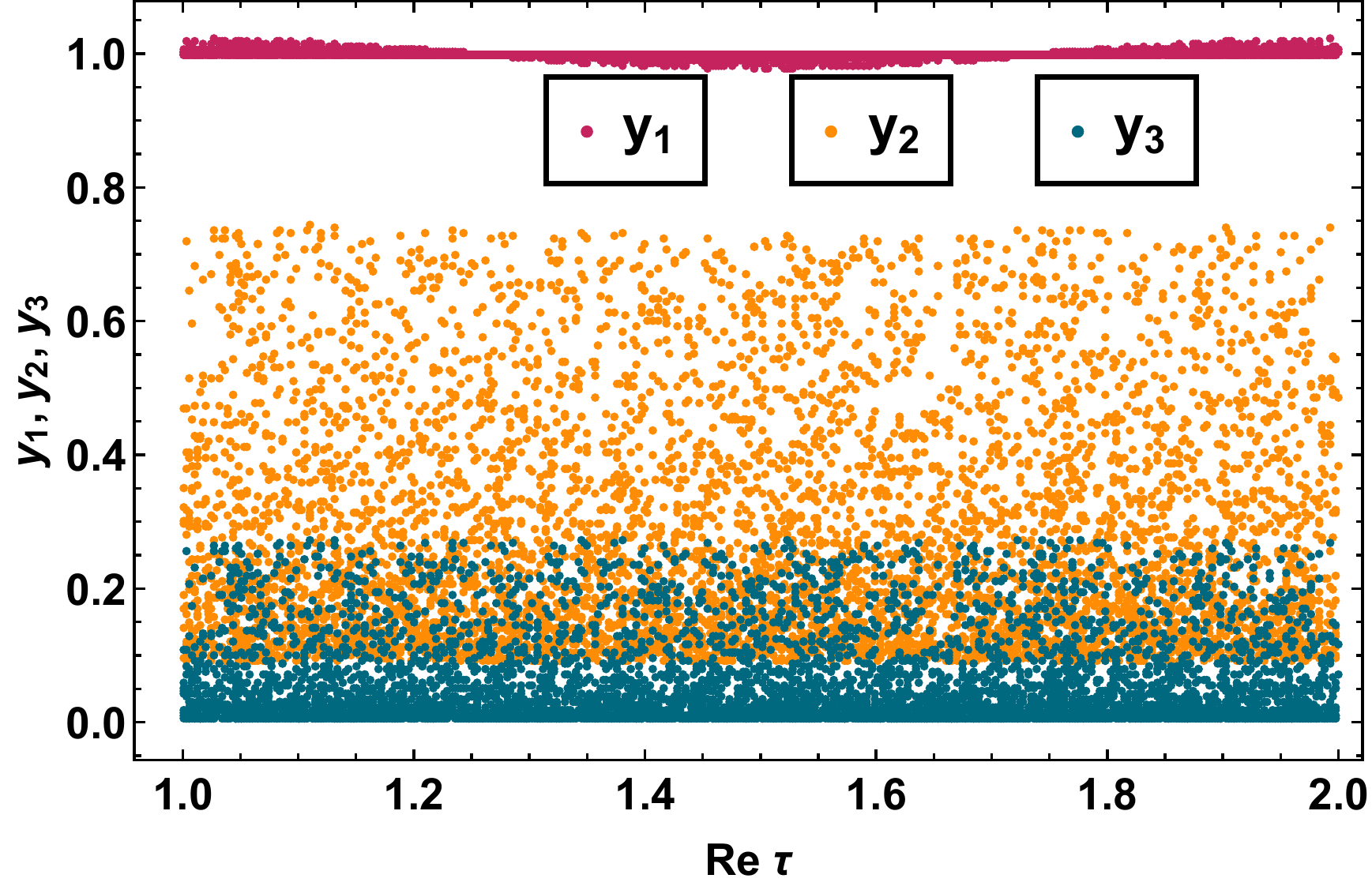}
\hspace*{0.2 true cm}
\includegraphics[height=50mm,width=75mm]{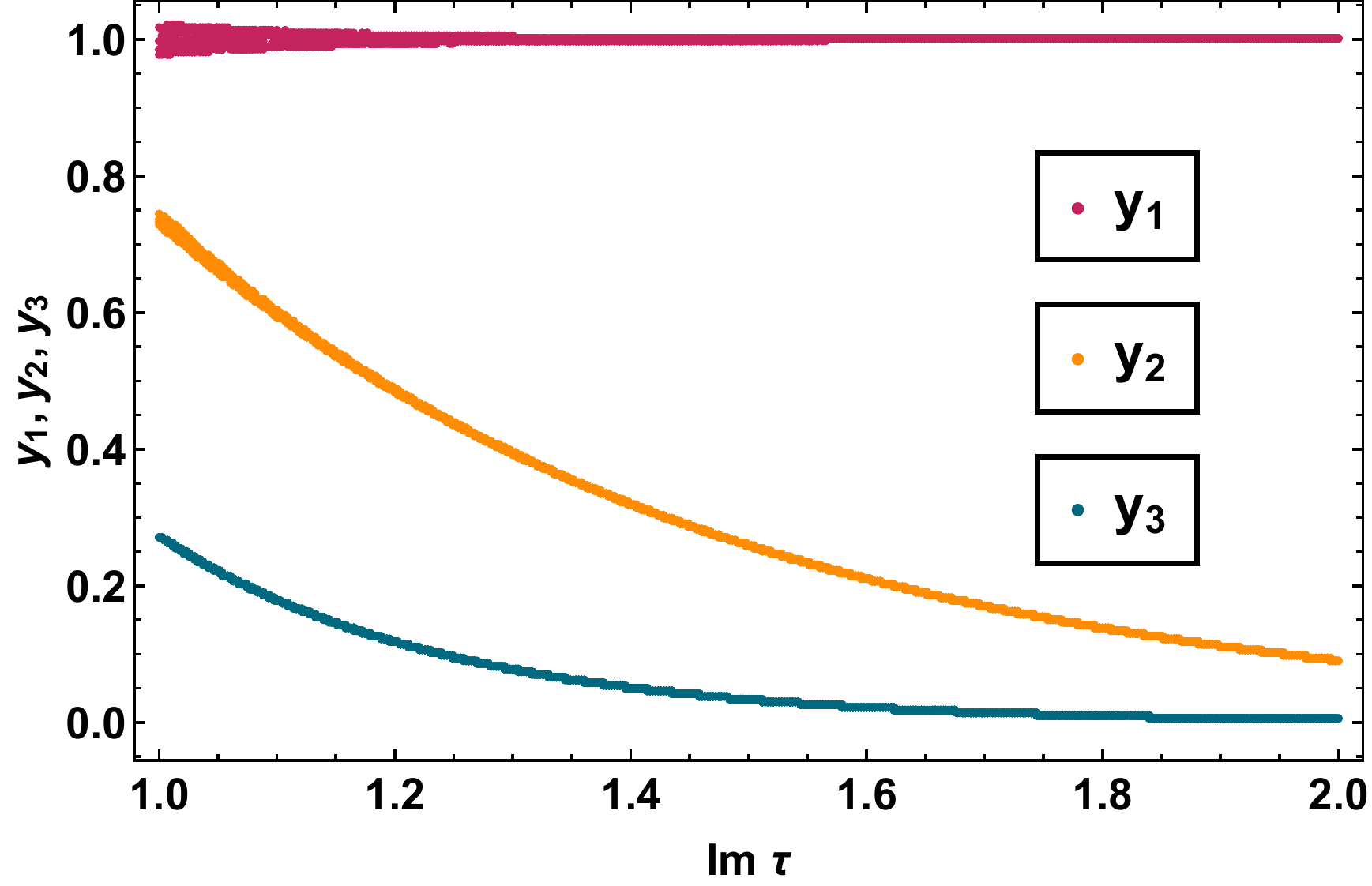}
\caption{Left panel indicates the interdependence of the modular Yukawa couplings ($y_1,y_2,y_3$) with the real part while right panel presents the imaginary part of modulus $\tau$. }
\label{yuk_reim_tau}
\end{center}
\end{figure}

\begin{figure}[h!]
\begin{center}
\includegraphics[height=50mm,width=75mm]{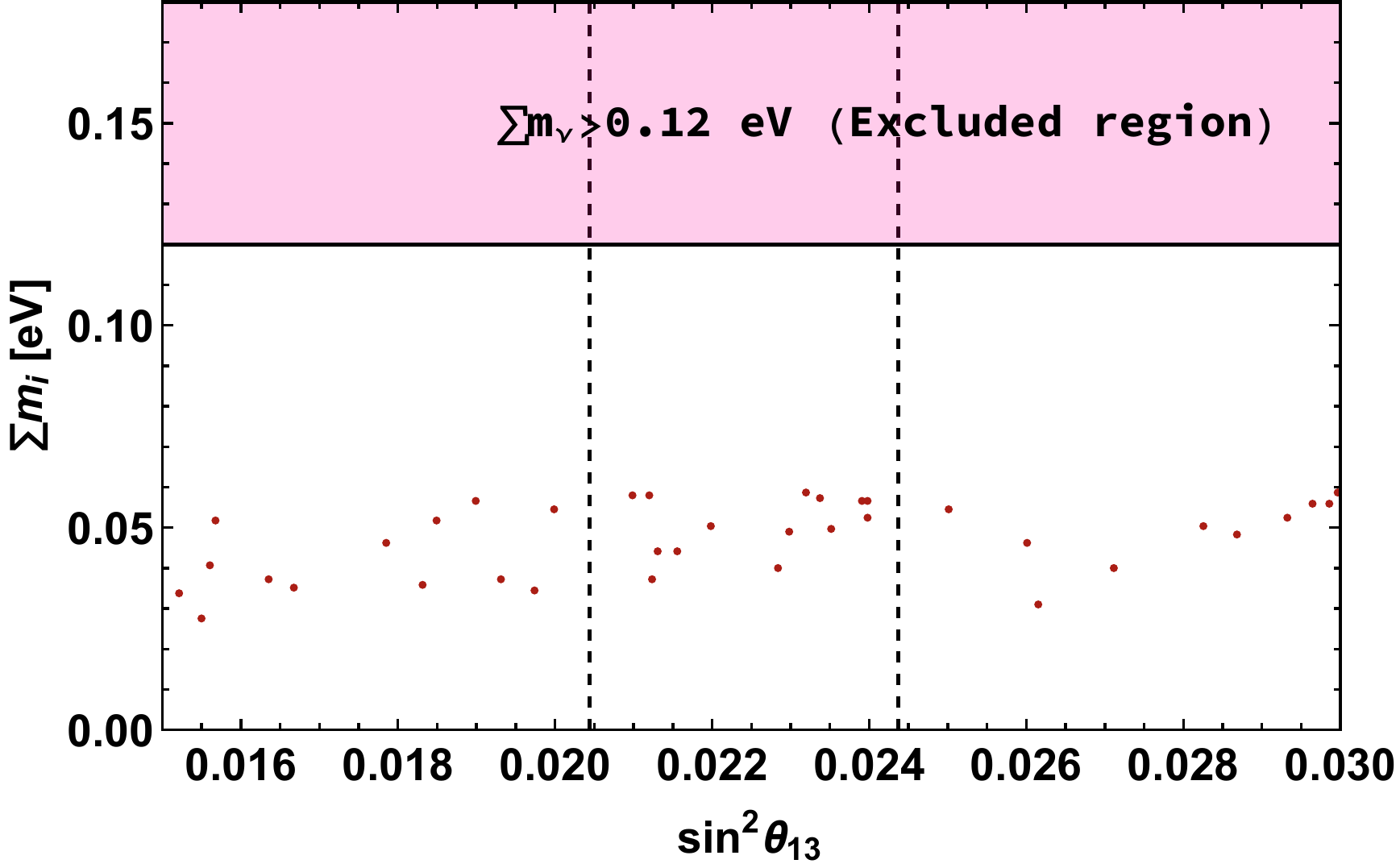}
\includegraphics[height=50mm,width=75mm]{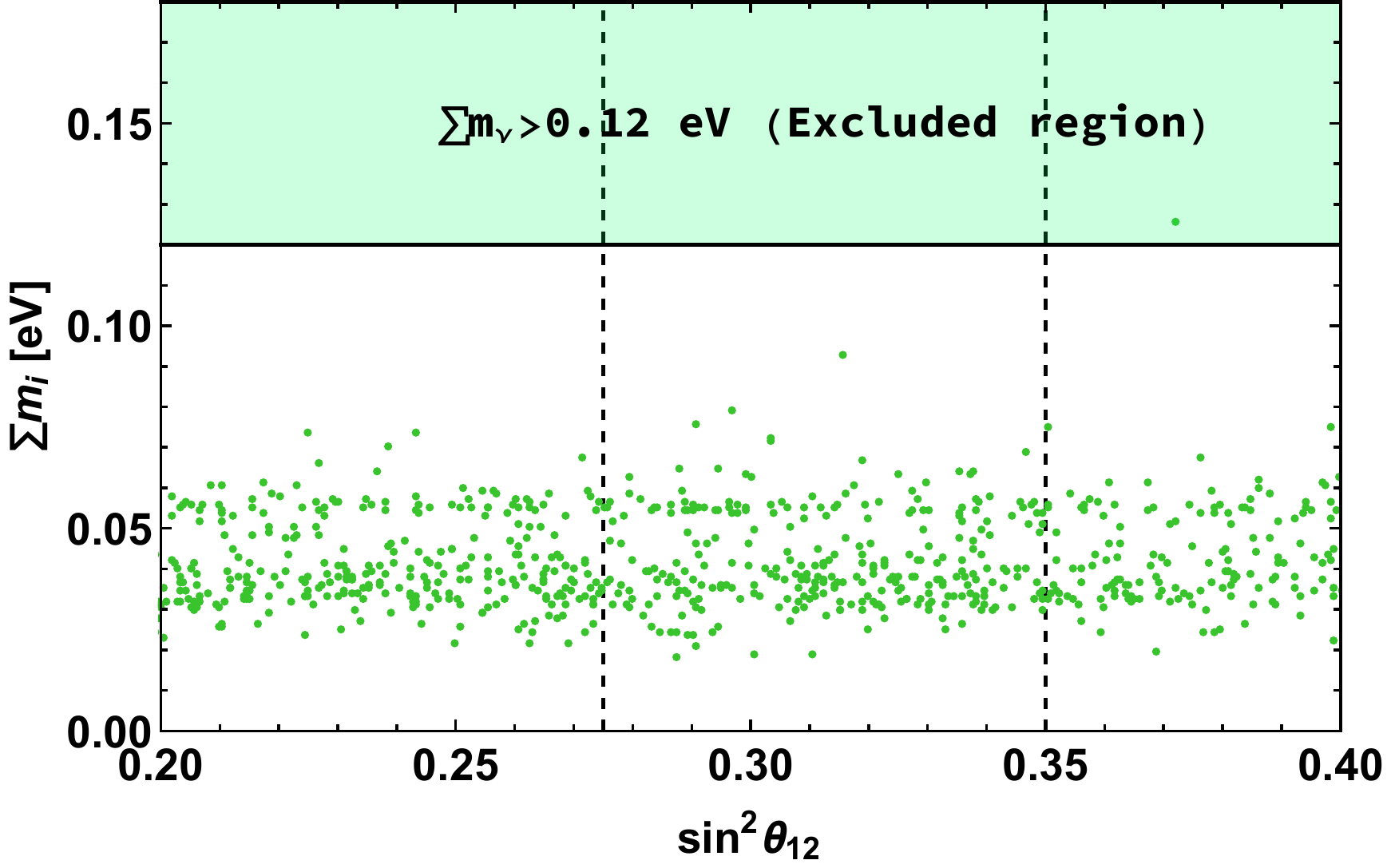}\\
\vspace*{0.3 true cm}
\includegraphics[height=50mm,width=75mm]{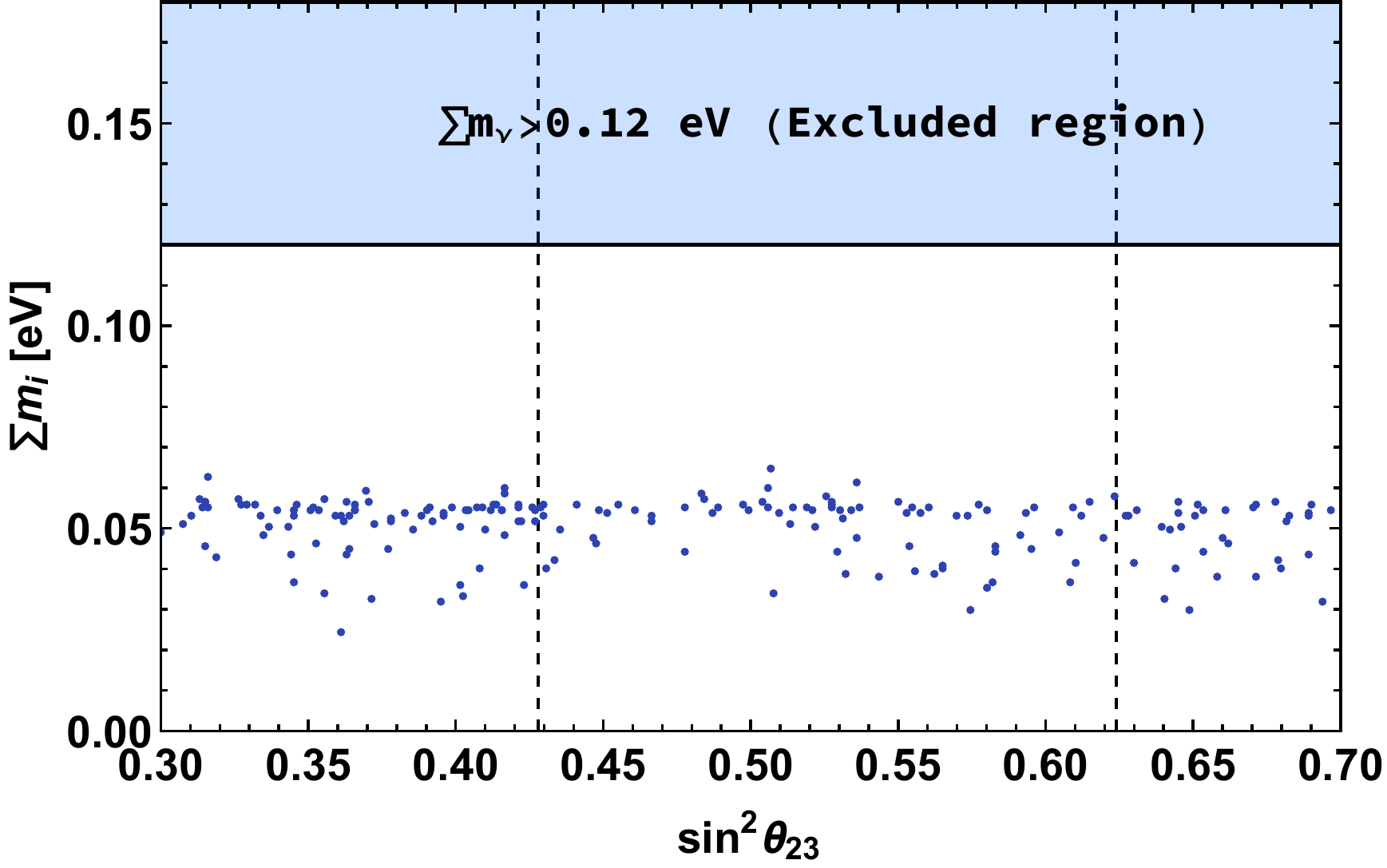}
\caption{Top left panel represents the interdependence of  $\Sigma m_i$ with $\sin^2 \theta_{13}$, and  $\sin^2\theta_{12}$ while the panel below displays its dependence on $\sin^2 \theta_{23}$.}
\label{mix_angles}
\end{center}
\end{figure}

\begin{figure}[h!]
\begin{center}
\includegraphics[height=50mm,width=75mm]{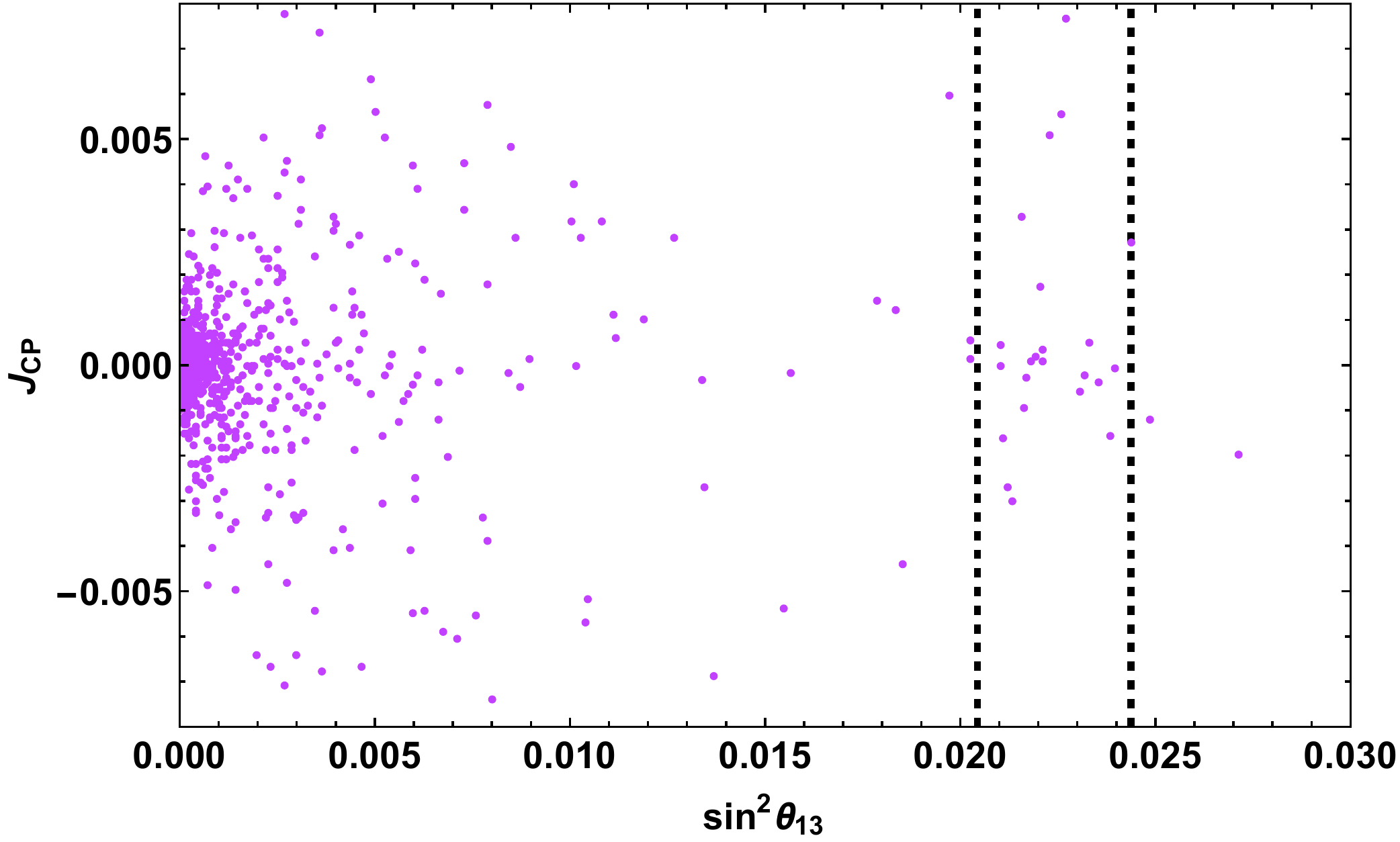}
\hspace*{0.2 true cm}
\includegraphics[height=50mm,width=75mm]{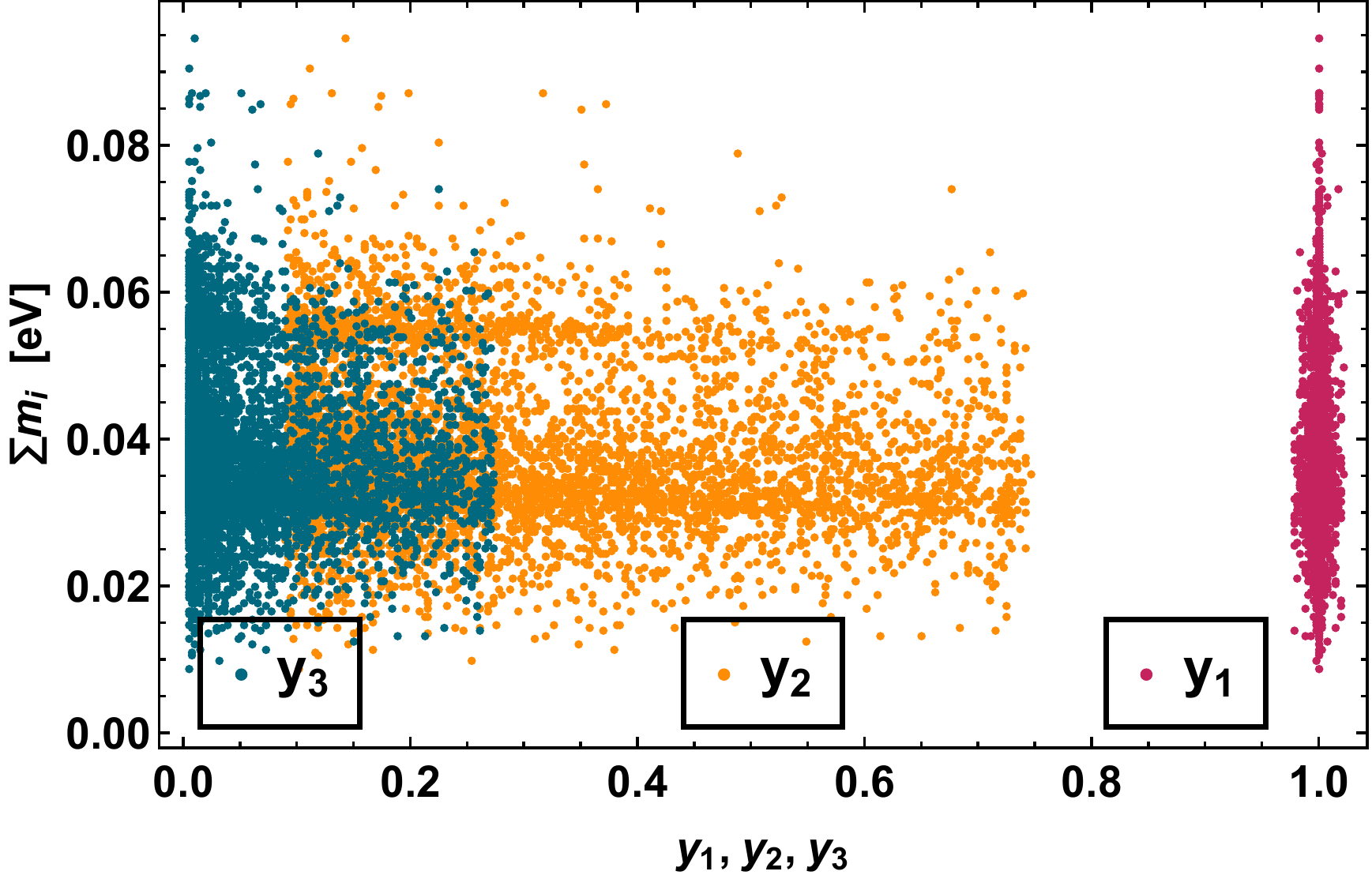}
\caption{Left panel makes an interdependence relation between the Jarlskog invariant with the reactor mixing angle while right panel reflects the alteration of sum of active neutrino masses  with the modular Yukawa couplings.}
\label{y_jcp}
\end{center}
\end{figure}

\begin{figure}[h!]
\begin{center}
\includegraphics[height=50mm,width=75mm]{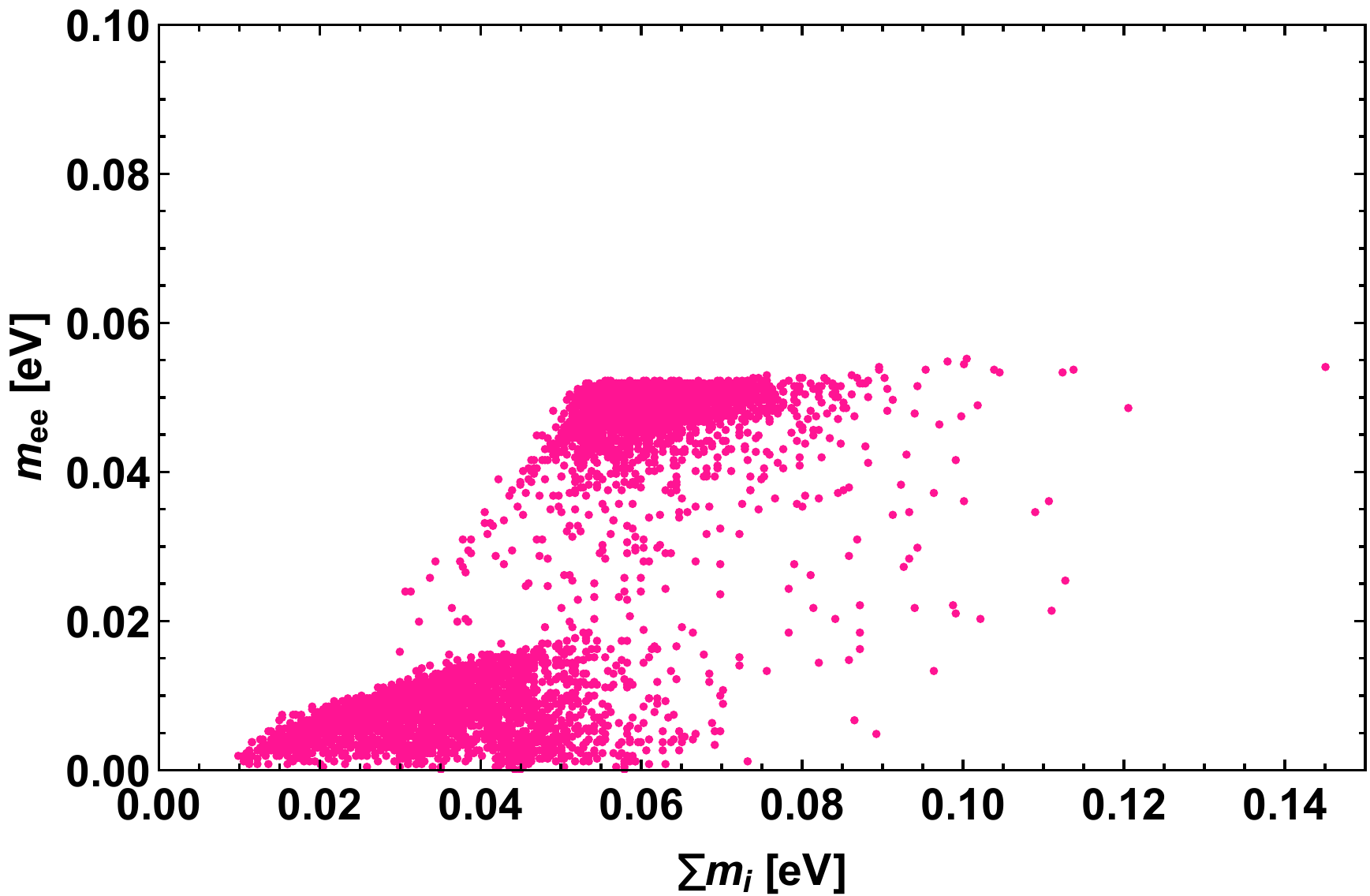}
\hspace*{0.2 true cm}
\includegraphics[height=48mm,width=72mm]{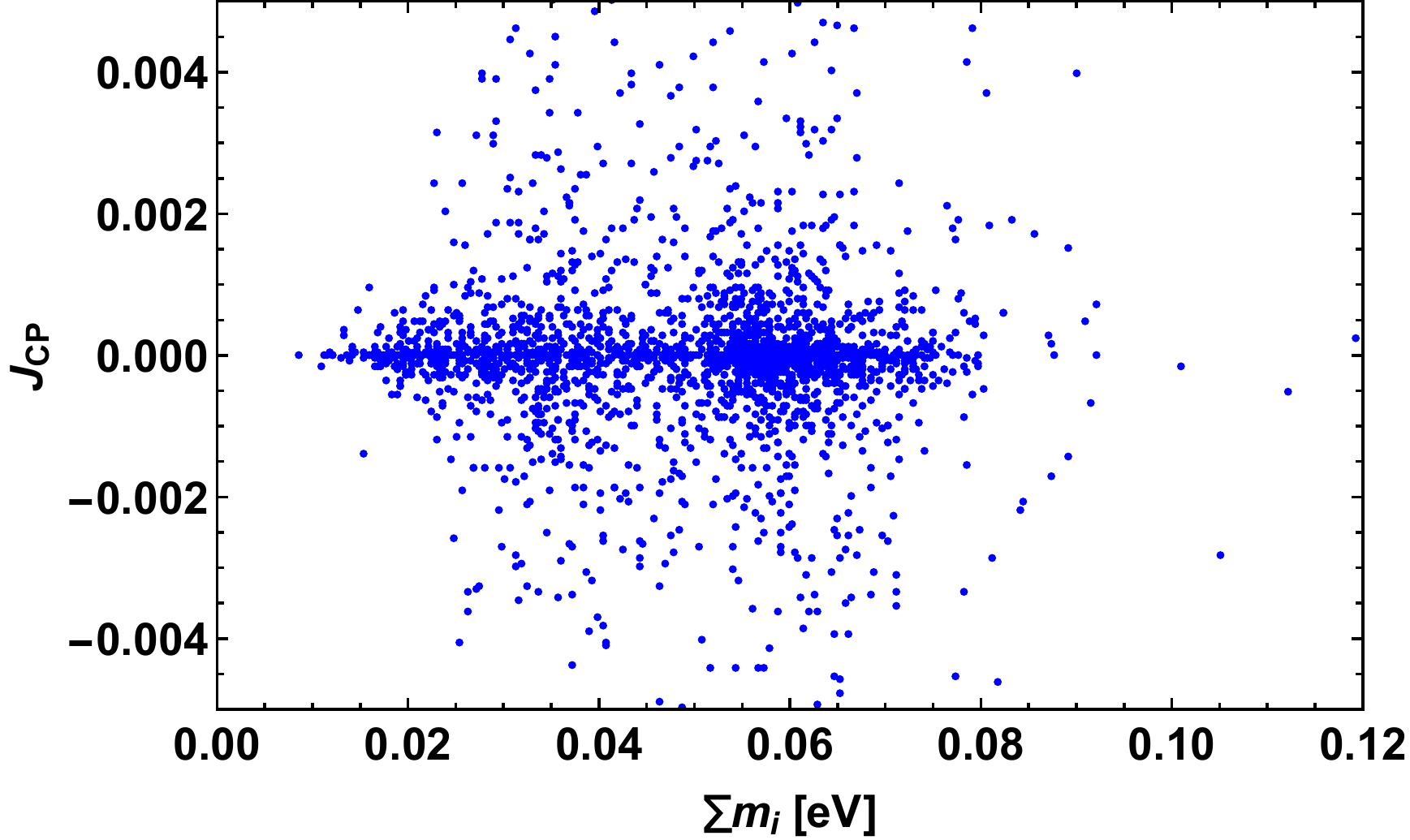}
\caption{Left panel above depicts the interdependence of effective neutrino mass of neutrinoless double beta decay with the sum of active neutrino masses, while, right panel shows the  relation of Jarlskog invariant with sum of active neutrino masses.}
\label{M23_mee}
\end{center}
\end{figure}

\begin{figure}[h!]
\begin{center}
\includegraphics[height=50mm,width=75mm]{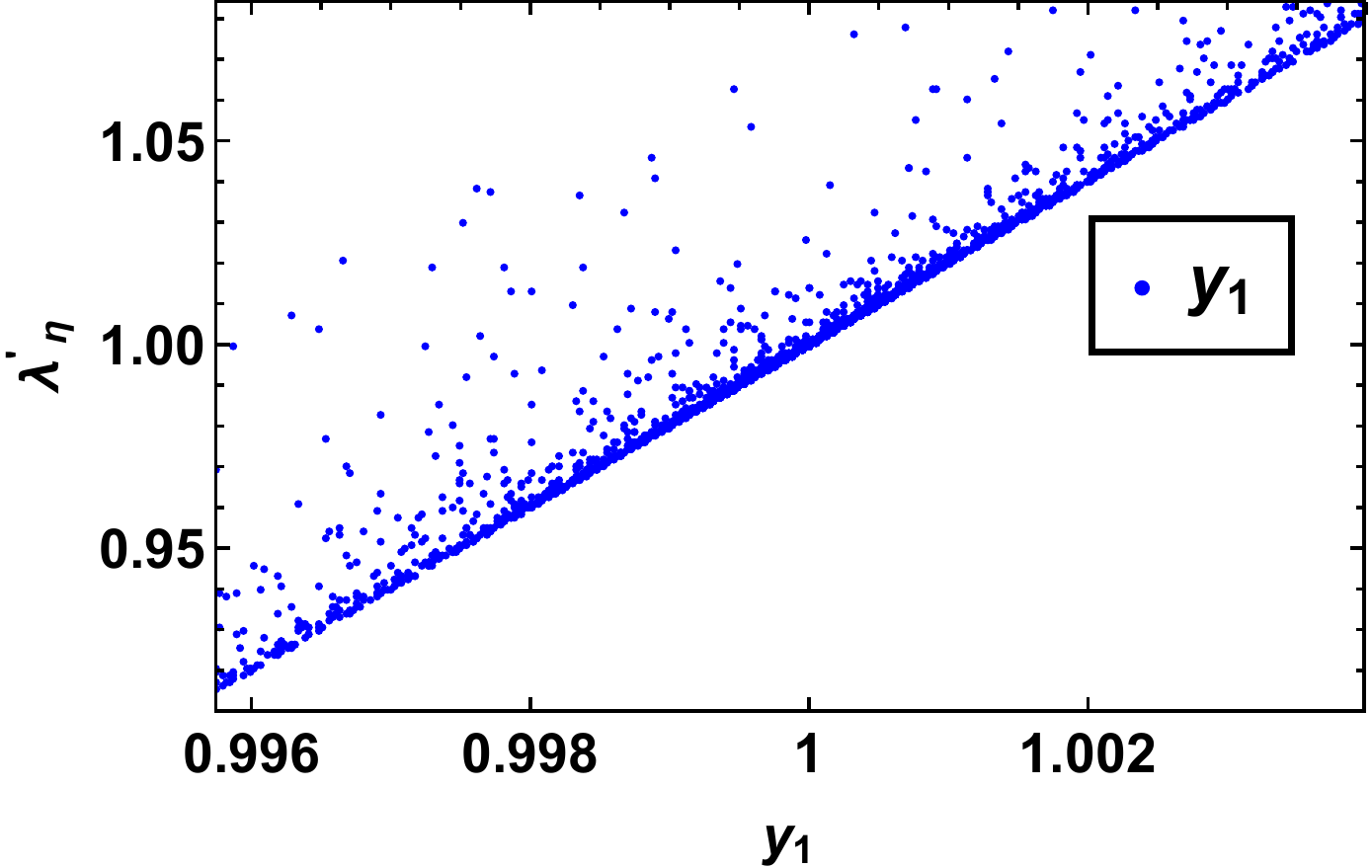}
\hspace*{0.2 true cm}
\includegraphics[height=50mm,width=75mm]{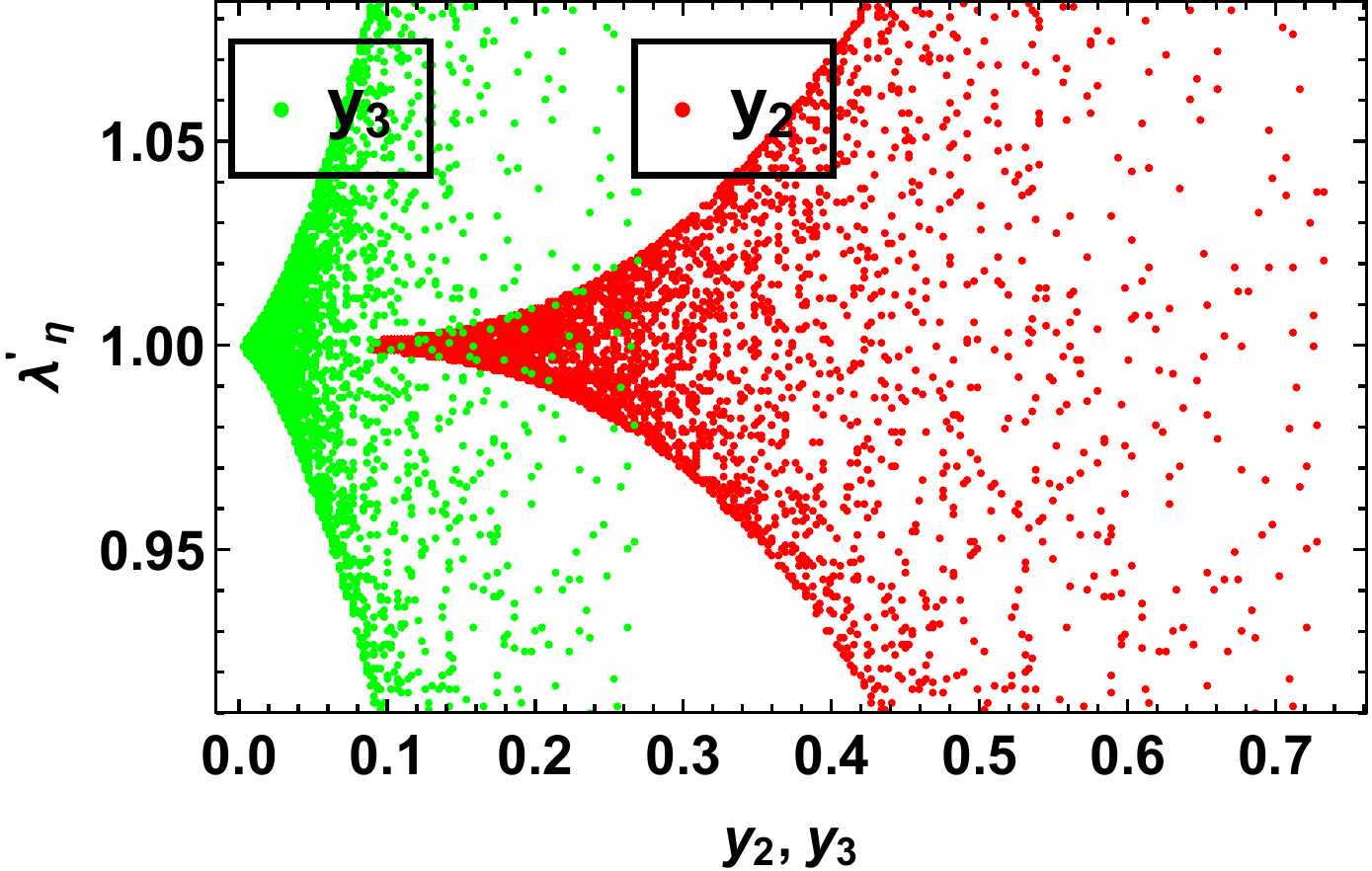}
\caption{Left (Right) panel displays the correlation between $\lambda_\eta^\prime$, which is an $A_4$ singlet and having a modular weight of $k=4$, with  $y_1$  ($y_2$,$y_3$).}
\label{lambda_y123}
\end{center}
\end{figure}

For constraining the model parameters, we use the current $3\sigma$ limit on neutrino mixing parameters for normal ordering (NO) from global-fit \cite{deSalas:2020pgw,Gariazzo:2018pei,Esteban:2020cvm}, which are given as
\bea
&&\Delta m^2_{\rm atm}=[2.431, 2.622]\times 10^{-3}\ {\rm eV}^2,~~~~~
\Delta m^2_{\rm sol}=[6.79, 8.01]\times 10^{-5}\ {\rm eV}^2, \nonumber\\ 
&&\sin^2\theta_{13}=[0.02044, 0.02437],\ ~
\sin^2\theta_{23}=[0.428, 0.624],\ ~
\sin^2\theta_{12}=[0.275, 0.350].  \label{eq:mix}
\eea
The model parameters are so chosen,  as to fit  the current neutrino oscillation data given in Eqn. (\ref{eq:mix}), as follows:
\begin{align}
&{\rm Re}[\tau] \in [1,2],~~{\rm Im}[\tau]\in [1,2],~~ \{ \alpha_{D},\beta_{D},\gamma_D \} \in [0.1,1.0],~~\{ \alpha^\prime_{D},\beta^\prime_{D},\gamma^\prime_D \} \in ~[0.1,1.0], \nn \\
& \quad \alpha_{NS}  \in [0.1, 0.5],\quad \beta_{NS} \in[0.05, 0.1],\quad v_\rho \in \nn [10^3,10^4] \ {\rm GeV},  \quad \Lambda \in  [10^4,10^5] \ {\rm GeV}.
\end{align}
The parameters used are randomly looked over the above mentioned ranges and the allowed regions for those are  first constrained  by the observed $3\sigma$ range of solar and atmospheric mass squared differences and further restricted by the observed sum of active neutrino masses  $\sum_i m_i < 0.12$ eV \cite{Aghanim:2019ame,Aghanim:2018eyx}. Furthermore, the range of modulus $\tau$ helps in validating the model with experimental results of neutrino masses (NO) is found to be 1\ $\lesssim\ $Re$[\tau]\lesssim$\ 2 and  1\ $\lesssim\ $Im$[\tau]\lesssim$\ 2. Hence, a very narrow range is satisfied by the modular Yukawa couplings, which are  functions of $\tau$ (please refer Appendix of \cite{Feruglio:2017spp}) and their regions of validation are found as:  0.99\ $\lesssim\ $$y_1$$(\tau)\lesssim$\ 1,  0.1\ $\lesssim\ $$y_2$$(\tau)\lesssim$\ 0.75 and 0.1\ $\lesssim\ $$y_3$$(\tau)\lesssim$\ 0.25. The behaviour  of  Yukawa couplings with respect to real and imaginary parts of $\tau$ are illustrated in the left and right panels of Fig. \ref{yuk_reim_tau} respectively. Proceeding further, Fig. \ref{mix_angles} depicts the alteration of the sum of total neutrino masses with the mixing angles  abiding to the $3\sigma$ regions. As mentioned in Sec. \ref{sec:radiative}, Fig. \ref{y_jcp}, helps us to have a glimpse of how Jarlskog CP invariant fits in the whole scenario, and found to be of the order of ${\cal O}(10^{-2})$, its connection with the reactor mixing angle is depicted in the left panel. The right panel of Fig. \ref{y_jcp}, expresses the complete parameter space for Yukawa couplings abiding to the sum of active neutrino masses. 
 Advancing further, the effective neutrino-less double beta decay mass parameter  $m_{ee}$ is found to have its value as $0.06$ eV as seen from the  left panel of Fig. \ref{M23_mee}, and the right panel of Fig. \ref{M23_mee}  shows the interdependence of Jarlskog invariant with sum of active neutrino mass.  As, here we are using a $A_4$ singlet coupling with $k_I=4$ ($\lambda_\eta'$), expressed in terms of Yukawa couplings i.e.,  triplet under $A_4$ with $k_I=2$, hence we explicitly show its correlation i.e.,  $\lambda^\prime_{\eta}$ with $y_1$ (left panel) and $\lambda^\prime_{\eta}$ with $y_2, y_3$ (right panel) of Fig.\ref{lambda_y123}.


\section{Comment on LFV Decay ($\mu \rightarrow e \gamma$) and muon {\lowercase{$g-2$}} anomaly}
\label{sec:comment}
\begin{figure}[h!]
\includegraphics[height=45mm,width=70mm]{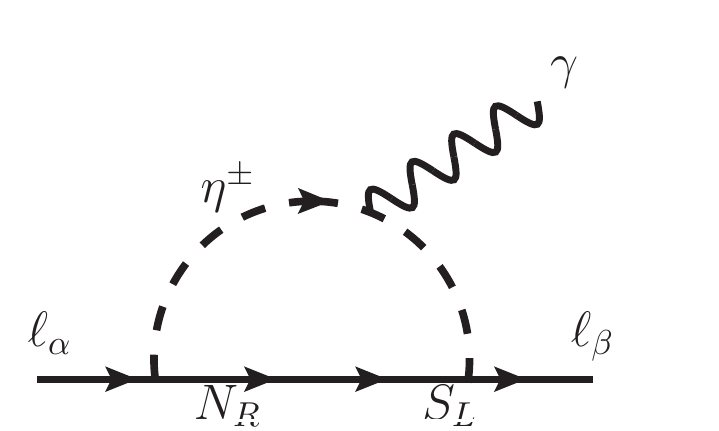}
\caption{Feynman diagram expressed here showcase LFV rare decays $\ell_\alpha \to \ell_\beta \gamma$  and muon $ g-2$  ($\alpha = \beta =\mu$) in context of current model.}\label{lfvfeyn}
\end{figure}
The quest in looking for lepton flavour violating decay mode $\mu \to e \gamma$ plays an exceptionally pivotal role in the hunt for new physics beyond the SM.
Many experiments are looking for this decay mode  with great effort for an improved sensitivity, and the current  limit on its branching   Br$(\mu\rightarrow e\gamma)< 4.2\times 10^{-13}$ is from MEG collaboration \cite{TheMEG:2016wtm}. Also the measured muon anomalous magnetic moment  shows around $3 \sigma$ discrepancy  with its SM predicted value, which is given as \cite{Tanabashi:2018oca,Dev:2020drf,Blum:2013xva,Bennett:2006fi}
\begin{equation}
\Delta a_\mu = a^{\rm exp}_\mu - a^{\rm SM}_\mu = (26.1 \pm 7.9)\times 10^{-10}.
\end{equation}
In the present framework,  the LFV process  $\mu \rightarrow e \gamma$ and muon $g-2$ occur at one loop level through standard Yukawa interactions. The Feynman diagram for this is displayed in Fig. \ref{lfvfeyn}. The branching ratio for the rare decay $\ell_\alpha \to \ell_\beta \gamma$ is given as  \cite{Chekkal:2017eka}
\begin{equation}
{\rm Br}(\ell_\alpha \rightarrow \ell_\beta \gamma)=\frac{3(4 \pi)^3 \alpha}{4 G_F^2}|A_D|^2\times {\rm Br}(\ell_\alpha \rightarrow \ell_\beta \nu_\alpha \bar{\nu}_{\beta}),
\end{equation}
where, $G_F\approx 10^{-5}~{\rm GeV}^{-2}$ (i.e. Fermi constant) and $\alpha$ being the  electromagnetic fine structure constant and $A_D$ is the dipole contribution, hence, expressed as
\begin{equation}
A_D=\sum_i \frac{(Y_D)_{\alpha i}~(Y^*_{LS})_{\beta i}~g(x)}{2(4\pi)^2 M^2_{\eta^\pm}}. \label{yukawaN}
\end{equation} 
Here, $Y_D$ and $Y_{LS}$ being the Yukawa coupling matrices as shown in eqn.\eqref{Eq:MD} and \eqref{Eq:Mls}, $g(x)$ is the loop function, with $x=\frac{M^2_{k}}{M^2_{\eta^\pm}}$, expressed as
\begin{equation}
g(x)=\frac{1}{6}\left[\frac{1-2x(3+1.5x+x^2-3x {\rm log}x)}{(1-x)^4}\right].
\end{equation} 
For $\alpha=\beta$,  the Feynman diagram of  Fig. \ref{lfvfeyn} will give contribution towards the muon anomalous magnetic moment, given as
\begin{equation}
\Delta a_\mu=\frac{1}{16 \pi^2}\left[\frac{m^2_\mu}{M^2_{\eta^\pm}}\sum_i (Y_D)_{\mu \mu} (Y^*_{LS})_{\mu \mu}~g(x)\right].
\end{equation}
The muon $g-2$ can also be obtained from the $Z^\prime$ and $\mu$ mediated loop, which will be suppressed due to the large mass difference.

\begin{figure}[h!]
\begin{center}
\includegraphics[height=50mm,width=75mm]{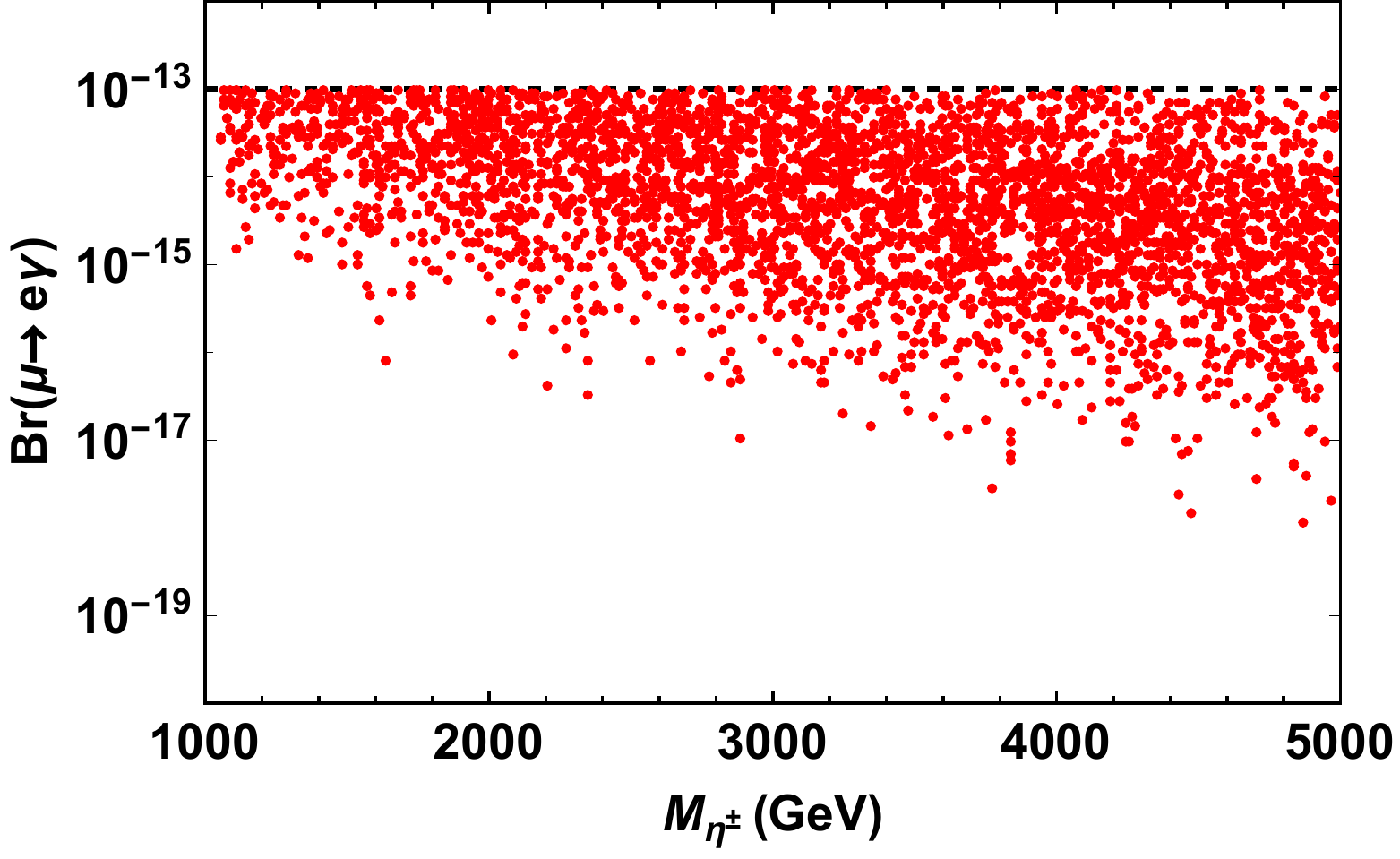}
\includegraphics[height=50mm,width=75mm]{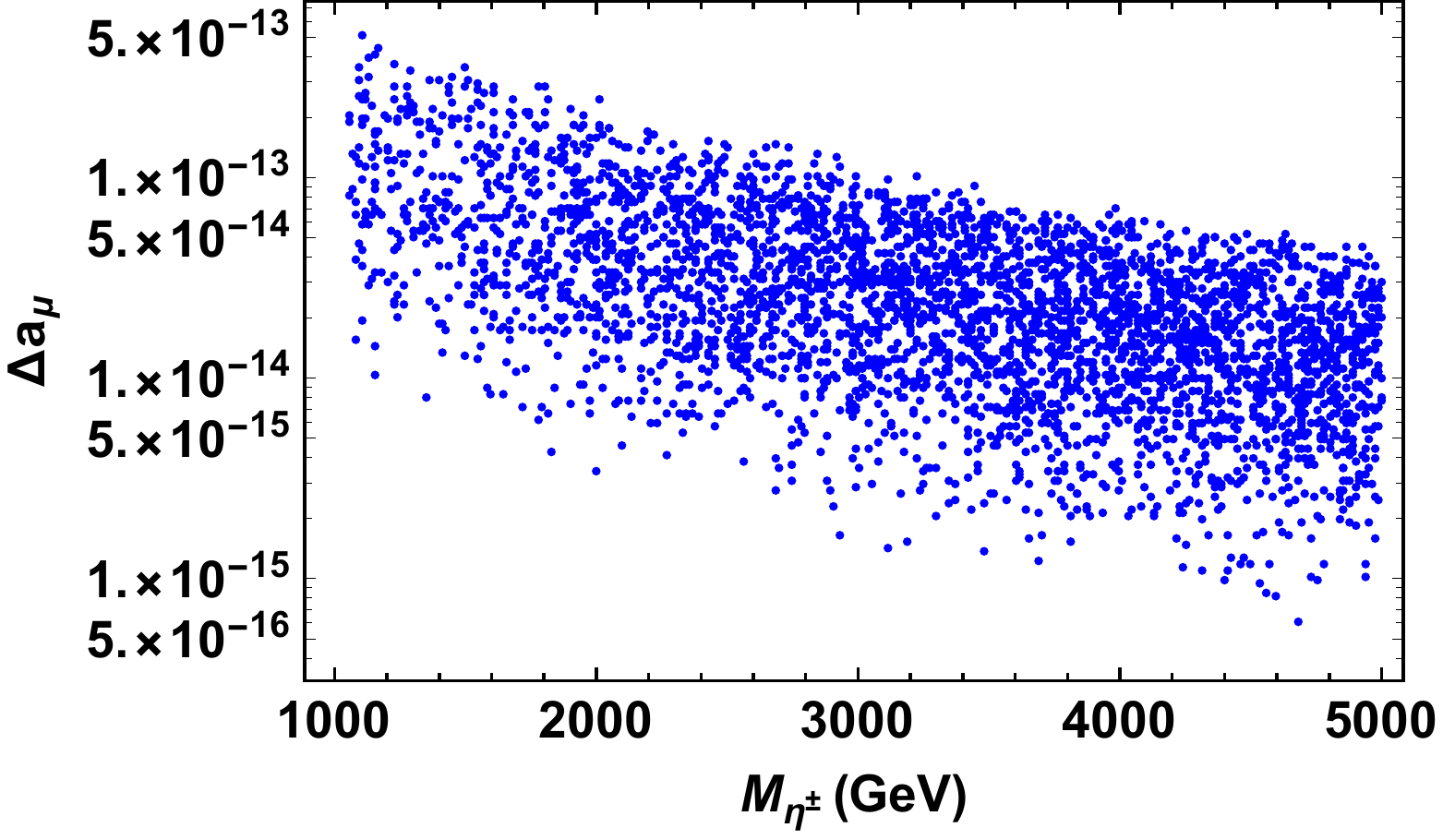}
\caption{The left (right) panel represents the variation of the LFV branching ratio of $\mu \rightarrow e \gamma$ process (muon $g-2$) with the charged inert scalar mass. }
\label{lfv1}
\end{center}

\end{figure}
\begin{figure}[h!]
\begin{center}
\includegraphics[height=50mm,width=75mm]{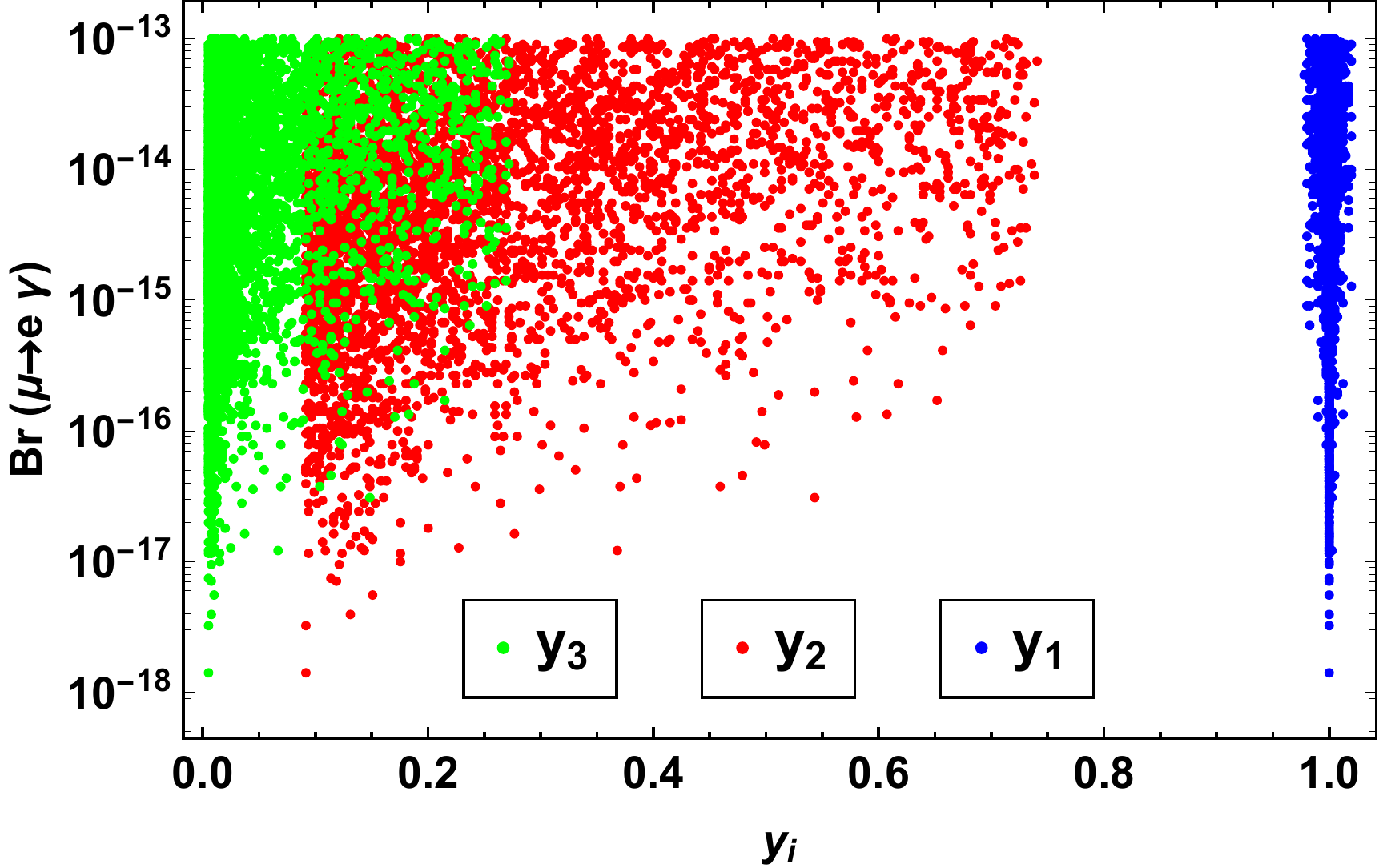}
\includegraphics[height=50mm,width=75mm]{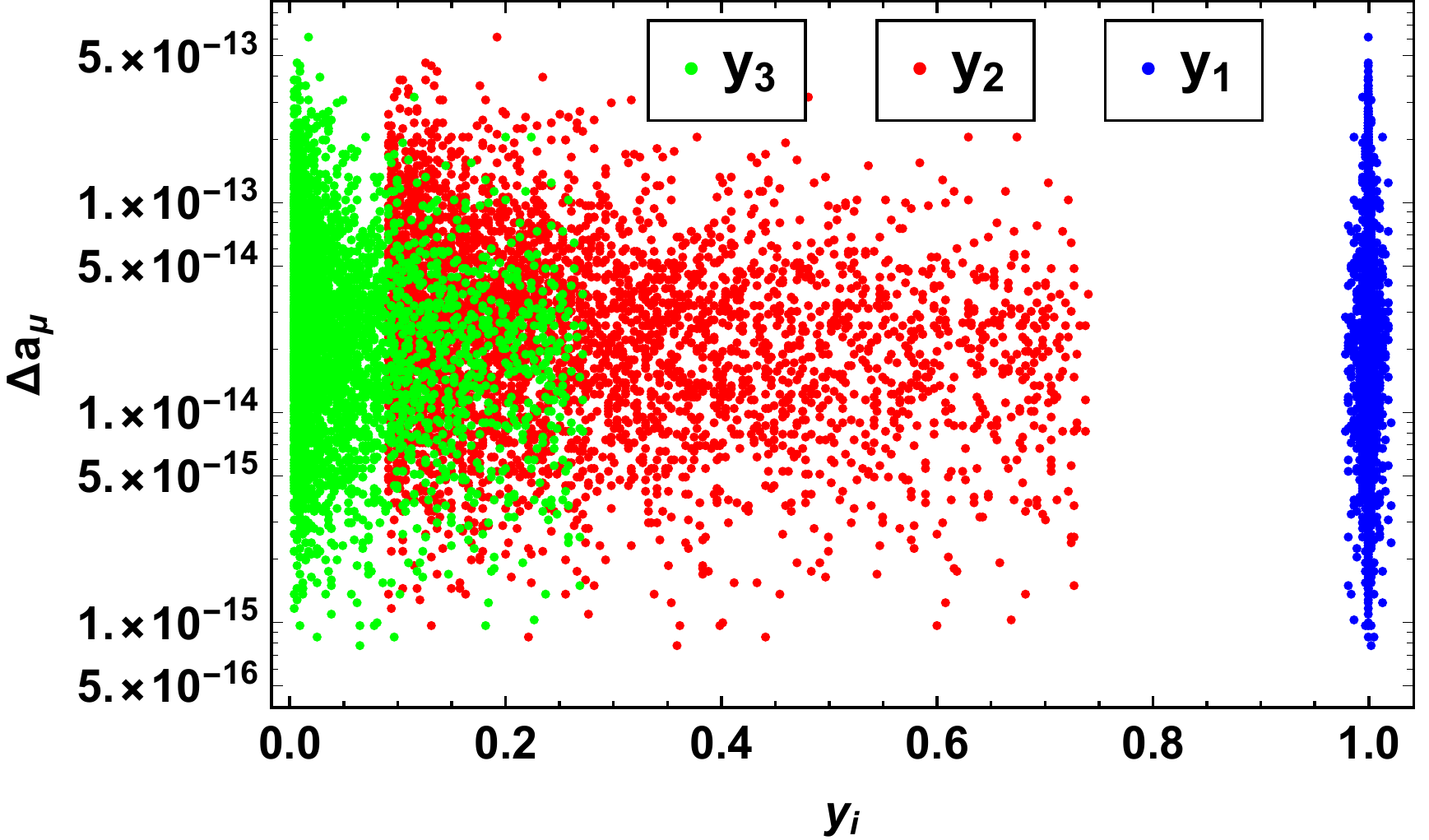}
\caption{Variation of the $\mu \to e \gamma$  branching fraction and muon $g-2$  with the Yukawa couplings  exhibited in the left and right panels respectively. }\label{Yuk_lfv}
\end{center}

\end{figure}

In the  left and right panels of Fig. \ref{lfv1}, we have represented the dependence of the branching fraction of $\mu \rightarrow e \gamma$ and anomalous muon magnetic moment $\Delta a_\mu$, on the inert charged  scalar mass,   which are found to lie within the experimental limits. 
The variation of  $\mu \to e \gamma $ branching fraction and $\Delta a_\mu$ with the modular Yukawa couplings, consistent with   neutrino mass constraints are displayed in Fig. \ref{Yuk_lfv}. 

\section{Fermionic Dark matter}
\label{sec:dark}

The model includes six heavy Majorana neutrinos which are doubly degenerate, out of which two of the lightest mass eigenstates can serve as dark matter candidates, provided the inert scalar particles are heavier. 

Before we move on to DM study, we first diagonalize the Majorana mass matrix of eqn. \ref{MRS_matrix}. For simplicity, we assume the coupling of symmetric part is dominant ($\alpha_{NS} > \beta_{NS}$). We diagonalize the reduced mass matrix with a TBM rotation and then by the normalized eigenvector matrix  \cite{Behera:2020sfe}. We have implemented the model in LanHEP package \cite{Semenov:1996es} and then extracted the results from micrOMEGAs \cite{Pukhov:1999gg, Belanger:2006is, Belanger:2008sj} package. 

\begin{figure}[h!]
\begin{center}
\includegraphics[height=35mm,width=55mm]{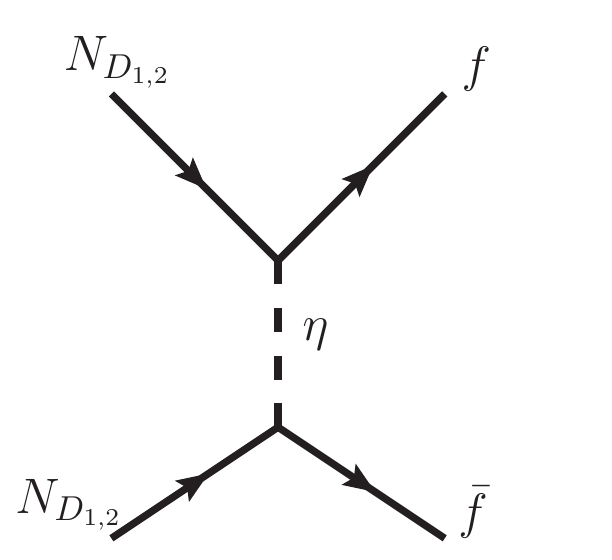}~~~~~~
\includegraphics[height=35mm,width=55mm]{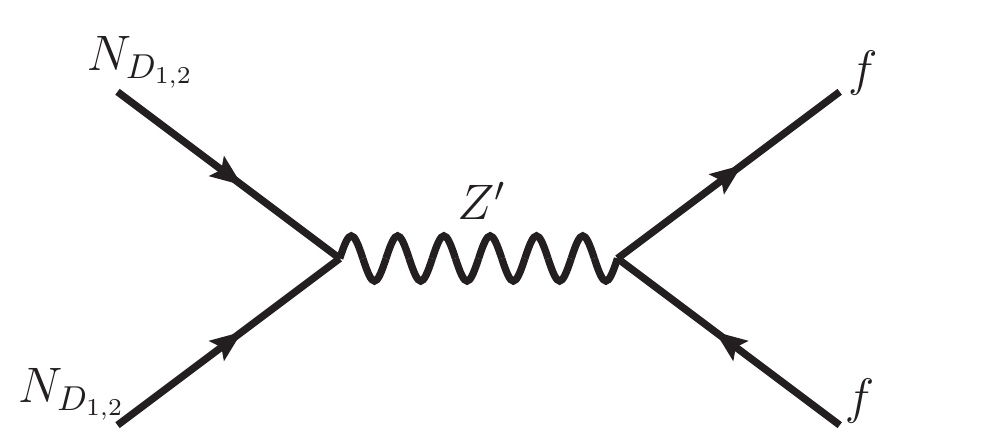}
\caption{Feynman diagrams for t and s-channel annihilation of DM, whose contribution is  towards the relic density.}
\label{relic_feyn}
\end{center}
\end{figure}

\begin{figure}[h!]
\begin{center}
\includegraphics[height=50mm,width=75mm]{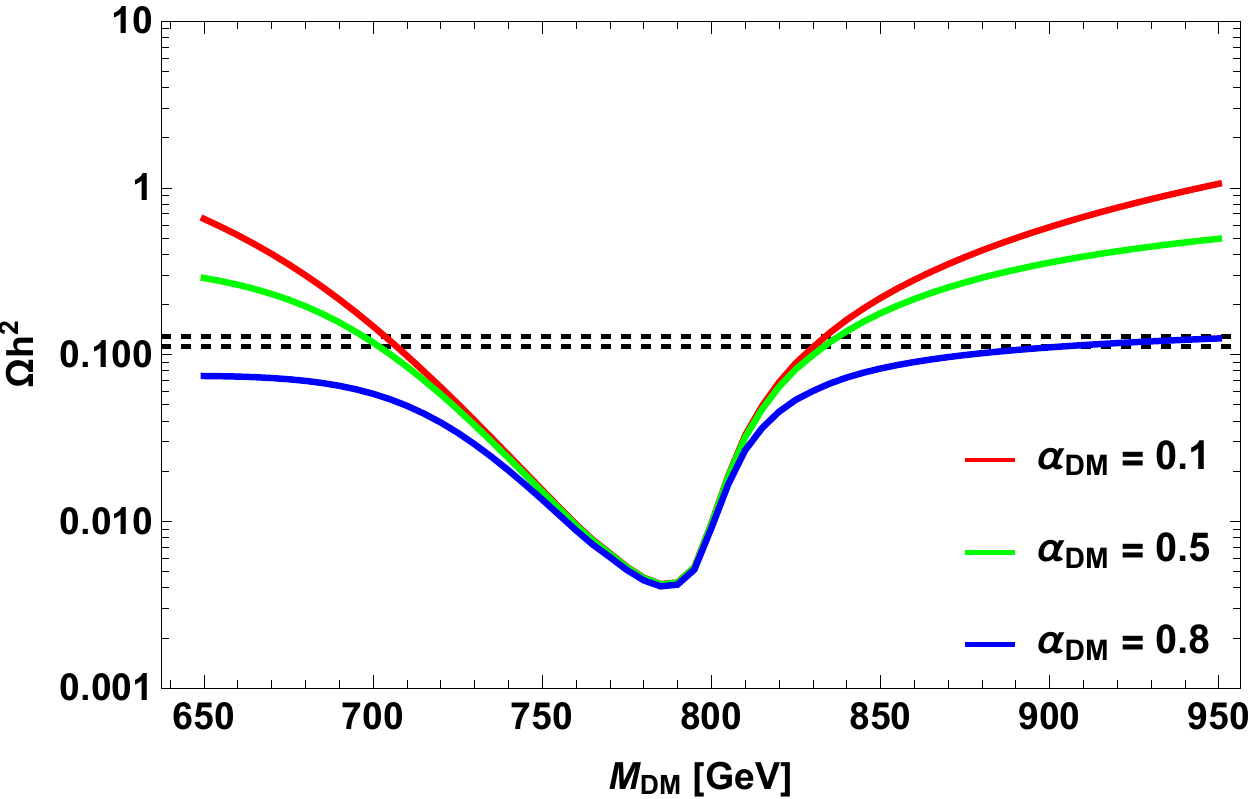}
\caption{Variation of abundance of fermionic DM as a function of its mass for various values of couplings. Black horizontal dashed lines stand for the $3\sigma$ bound of Planck satellite data \cite{Aghanim:2018eyx}.}
\label{relic_plot}
\end{center}
\end{figure}

We wish to compute the relic density for a particular benchmark. We confine our discussion by fixing $\alpha_{NS} = 0.5$, $v_{\rho} = 5$ TeV and also the Yukawa couplings in the range $0.1 \lesssim y_{2,3} \lesssim 0.25$. As we see from  Fig. \ref{yuk_reim_tau}, $y_1$ does not vary much and thus, $y_{2,3}$ dictate the mass range of DM i.e., $\sim 650 - 950$ GeV. Choosing equal values ($\alpha_{\rm DM}$) for the couplings $\alpha_D,\beta_D,\gamma_D$ and $\alpha^\prime_D,\beta^\prime_D,\gamma^\prime_D$, we project the DM abundance as a function of its mass in Fig. \ref{relic_plot}.  The annihilation channels (shown in Fig. \ref{relic_feyn}) with lepton and anti-lepton pair in the final state in $\eta$-portal ($t$-channel) and $Z^\prime$-portal ($s$-channel), contribute to relic density. One can see that  the $s$-channel contribution gives resonance on the either side of $M_{\rm DM} = M_{Z^\prime}/{2}$, with $M_{Z^\prime} = 1.6$ TeV. 

Moving to detection prospects, $\eta$ and $Z^\prime$  have no direct interactions with quarks, hence study of tree-level DM-nucleon scattering is not possible. One-loop contribution to DM scattering off nuclei will be well below experimental upper limits (both spin-independent and spin-dependent) and do not show any impact on model parameters \cite{Ibarra:2016dlb}.
 

\section{Conclusion}
\label{sec:con}
In this paper, the main motive of the model is to implement $A_4$ modular symmetry to see its novelty in neutrino phenomenology through scotogenic framework. We have realized neutrino mass at one loop level successfully by introducing an inert scalar doublet $\eta$ ($A_4$ singlet with modular weight $-2$) and six heavy fermions $N_R$ and $S_L$ (triplets under $A_4$ with modular weights $-1$ and $+1$ respectively). As we are dealing with $A_4$ modular symmetry, the Yukawa couplings are defined as $A_4$ triplet ($\bm Y$) with modular weight $2$, and the scalar couplings for terms involving $\eta$ as $A_4$ singlets ($\bm \lambda_\eta$,   $\bm \lambda_\eta^\prime$) with weights $4,8$ respectively.  An additional $U(1)_X$ is imposed to avoid unwanted Majorana mass terms and a complex scalar singlet $\rho$ is introduced to spontaneously break this local gauge symmetry. 

Modular symmetry not only avoids adding new flavon fields for neutrino phenomenology but also plays a vital role in ensuring dark matter stability. A particular flavor structure for the neutrino mass matrix is achieved along with neutrino mixing. We have used the procedure of numerically diagonalising the neutrino mass matrix and fixed the model parameters in such a way that they remain compatible with present $3\sigma$ range of oscillation data. Proceeding further, we have established the present model's contribution towards lepton flavor violating decay $\mu \to e\gamma$, compatible with upper bound set by MEG collaboration. We also found that the contribution to muon $g-2$ anomaly (i.e. $\Delta a_\mu$) is in the range of  $10^{-12} - 10^{-14}$ satisfying the experimental cut-off. Finally, we have addressed dark matter phenomenology of the lightest stable fermion spectrum. With stringent bounds on Yukawa couplings confining dark matter mass, we have obtained the relic density compatible to Planck data for a particular benchmark of values for model parameters. We found that the annihilations with lepton-anti lepton pair in the final state  via $\eta$ and $Z^\prime$ ($U(1)_X$ associated) portal contribute to relic density. Tree-level direct detection is not feasible as $\eta$ and $Z^\prime$ do not couple to quarks directly. To conclude, $A_4$ modular symmetry stands tall, providing rich neutrino phenomenology by avoiding the set of flavon fields as used in the conventional frameworks and also stabilizing dark matter candidate. The present paper remains an example, discussing the above aspects in the light of modular symmetry.

\acknowledgments

MKB and SM want  to acknowledge DST for its financial help. RM  acknowledges the support from  SERB, Government of India, through grant No. EMR/2017/001448.

\bibliographystyle{my-JHEP}
\bibliography{scoto}

\end{document}